\documentclass[twocolumn]{aastex62}
\usepackage{amsmath}
\hypersetup{
	colorlinks	= true,
	linkcolor	= red,
	urlcolor	= cyan,
	citecolor	= blue
}
\usepackage{amssymb}
\usepackage{lipsum}
\usepackage{multirow}

\usepackage[version=4]{mhchem}
\usepackage{ragged2e}
\usepackage{gensymb}
\usepackage{tikz}
\usepackage{graphics}
\usepackage[figure,figure*]{hypcap}
\usepackage[para,online,flushleft]{threeparttable}
\usepackage{comment}
\usepackage{tablefootnote}
\usepackage{subscript}

\makeatletter
\newcounter{reaction}
\renewcommand\thereaction{R\arabic{reaction}}
\@addtoreset{reaction}{chapter}
\newcommand\reactiontag%
{\refstepcounter{reaction}\tag{\thereaction}}
\newcommand\reaction@[2][]%
{\begin{equation}\ce{#2}%
\ifx\@empty#1\@empty\else\label{#1}\fi%
\reactiontag\end{equation}}
\newcommand\reaction@nonumber[1]%
{\begin{equation*}\ce{#1}\end{equation*}}
\newcommand\reaction%
{\@ifstar{\reaction@nonumber}{\reaction@}}
\makeatother

\graphicspath{{./images/}}

\turnoffediting

\shortauthors{Hu et al.}

\begin{document}
	
	\title{Unveiling shrouded oceans on temperate sub-Neptunes via transit signatures of solubility equilibria vs. gas thermochemistry}
	
	\correspondingauthor{Renyu Hu}
	\email{renyu.hu@jpl.nasa.gov \\ @2021 California Institute of Technology. \\ Government sponsorship acknowledged.}
	
	\author[0000-0003-2215-8485]{Renyu Hu}
	\affiliation{Jet Propulsion Laboratory, California Institute of Technology, Pasadena, CA 91109, USA}
	\affiliation{Division of Geological and Planetary Sciences, California Institute of Technology, Pasadena, CA 91125, USA}
	
	\author[0000-0002-1830-8260]{Mario Damiano}
	\affiliation{Jet Propulsion Laboratory, California Institute of Technology, Pasadena, CA 91109, USA}
	
	\author[0000-0003-4331-2277]{Markus Scheucher}
	\affiliation{Jet Propulsion Laboratory, California Institute of Technology, Pasadena, CA 91109, USA}
	\affiliation{Institut f\"ur Planetenforschung, Deutsches Zentrum f\"ur Luft- und Raumfahrt, D-12489 Berlin, Germany}
	
	\author[0000-0002-1426-1186]{Edwin Kite}
	\affiliation{Department of the Geophysical Sciences, University of Chicago, Chicago, IL 60637, USA}
	
	\author[0000-0002-6892-6948]{Sara Seager}
	\affiliation{Department of Earth, Atmospheric, and Planetary Sciences, Massachusetts Institute of Technology, Cambridge, MA 02139, USA}
	\affiliation{Department of Physics and Kavli Institute for Astrophysics and Space Research, Massachusetts Institute of Technology, Cambridge, MA 02139, USA}
	\affiliation{Department of Aeronautics and Astronautics, Massachusetts Institute of Technology, Cambridge, MA 02139, USA}
	
	\author{Heike Rauer}
	\affiliation{Institut f\"ur Planetenforschung, Deutsches Zentrum f\"ur Luft- und Raumfahrt, D-12489 Berlin, Germany}
	\affiliation{Institut f\"ur Geologische Wissenschaften, Freie Universit\"at Berlin, D-12249 Berlin, Germany}
	
	\begin{abstract}
The recent discovery and initial characterization of sub-Neptune-sized exoplanets that receive stellar irradiance of approximately Earth's raised the prospect of finding habitable planets in the coming decade, because some of these temperate planets may support liquid water oceans if they do not have massive H$_2$/He envelopes and are thus not too hot at the bottom of the envelopes. For planets larger than Earth, and especially planets in the $1.7-3.5\ R_{\oplus}$ population, the mass of the H$_2$/He envelope is typically not sufficiently constrained to assess the potential habitability. Here we show that the solubility equilibria vs. thermochemistry of carbon and nitrogen gases results in observable discriminators between small H$_2$ atmospheres vs. massive ones, because the condition to form a liquid-water ocean and that to achieve the thermochemical equilibrium are mutually exclusive. The dominant carbon and nitrogen gases are typically CH$_4$ and NH$_3$ due to thermochemical recycling in a massive atmosphere of a temperate planet, and those in a small atmosphere overlying a liquid-water ocean are most likely CO$_2$ and N$_2$, followed by CO and CH$_4$ produced photochemically. NH$_3$ is depleted in the small atmosphere by dissolution into the liquid-water ocean. These gases lead to distinctive features in the planet's transmission spectrum, and a moderate number of repeated transit observations with the James Webb Space Telescope should tell apart a small atmosphere vs. a massive one on planets like K2-18~b. This method thus provides a way to use near-term facilities to constrain the atmospheric mass and habitability of temperate sub-Neptune exoplanets.
	\end{abstract}
	
	
	\section{Introduction} \label{sec:intro}
	
The exoplanet community already has ways to detect an H$_2$ atmosphere by transmission spectroscopy via its pressure scale height one order of magnitude larger than that of an N$_2$ or CO$_2$ atmosphere \citep{miller2008atmospheric}. However, the mass of the H$_2$ atmosphere -- the parameter that controls the temperature at the bottom of the atmosphere and thus the possibility for liquid water \citep{pierrehumbert2011hydrogen,ramirez2017volcanic,koll2019hot} -- is not directly measurable from the transmission spectrum. Also, a planet's mass and radius typically allow multiple models of the interior structure \citep[e.g.,][]{rogers2010three,valencia2013bulk}. It is unclear whether the planets in the $1.7-3.5\ R_{\oplus}$ population \citep{fulton2018california} are mostly rocky planets with massive H$_2$/He gas envelopes \citep{owen2017evaporation,jin2018compositional} or planets with a massive water layer ($\sim50$ wt. \%) that do not require a large H$_2$ envelope to explain their radius \citep[e.g., referred to as ``ocean planets'' thereafter;][]{zeng2019growth,mousis2020irradiated,venturini2020nature}. Direct-imaging observations in the future may provide means to detect a surface underneath a thin atmosphere on temperate planets, via the ocean glint \citep{robinson2010detecting} or surface heterogeneity \citep{cowan2009alien,fan2019earth}. However, these methods are not applicable to the near-term capabilities such as the JWST and may pose challenges on precision even for ambitious direct-imaging mission concepts \citep{Gaudi2020}. 

\begin{figure}[!htbp]
\centering
\includegraphics[width=0.45\textwidth]{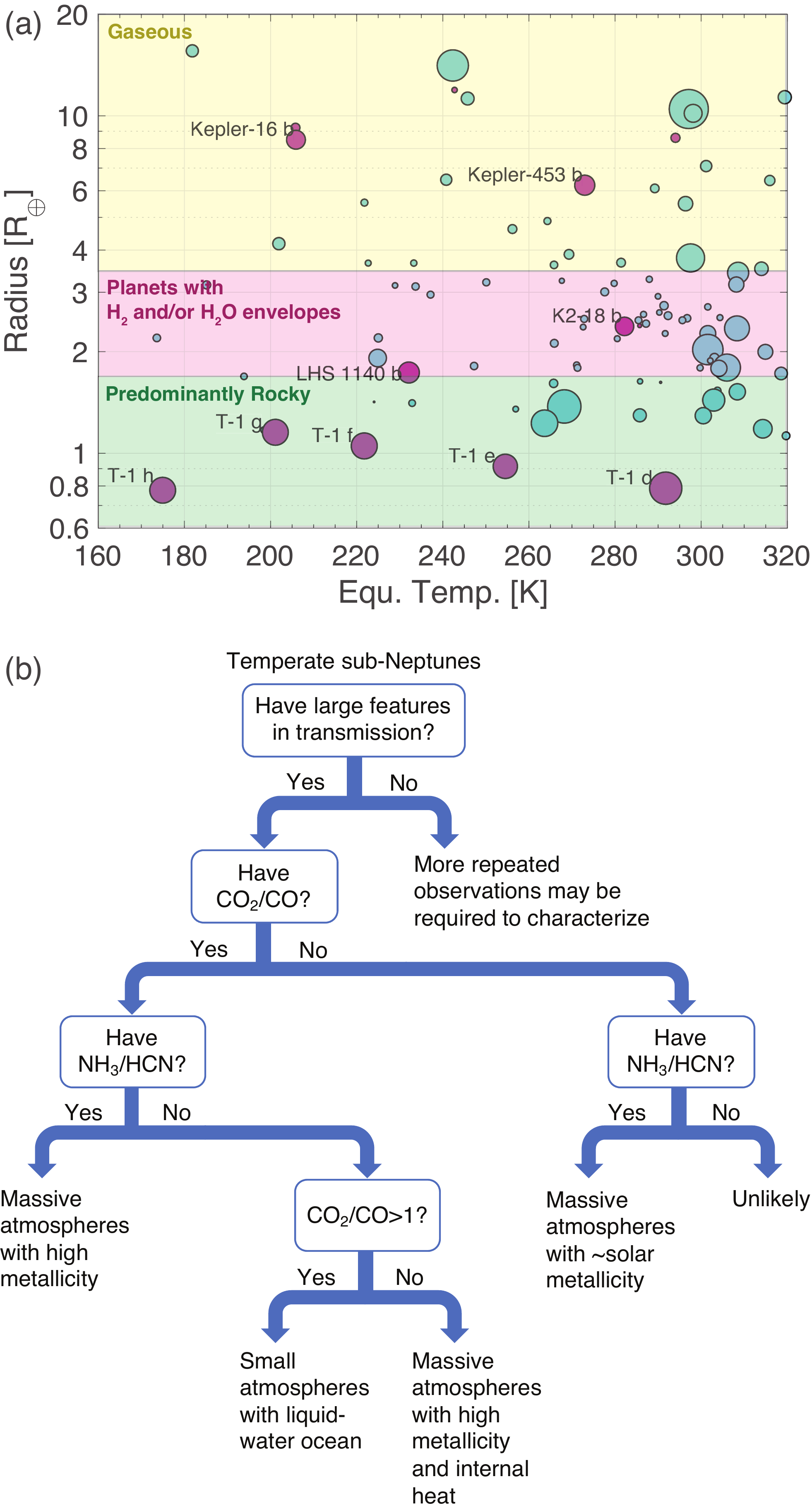}
\caption{
Temperate exoplanets amenable for atmospheric characterization via transmission spectroscopy. (a) Purple dots are confirmed planets with measured masses, and blue dots are planets with unknown masses or planet candidates. Data are taken from the NASA Exoplanet Archive and and the TESS Objects of Interest Catalog. The marker sizes are scaled with the expected S/N of the spectral features of an H$_2$ atmosphere observed by JWST at $2\ \mu m$. Most of the temperate planets and planet candidates suitable for atmospheric characterization are larger than Earth and thus more likely to have H$_2$ atmospheres. (b) A roadmap to characterize the mass of the atmospheres and the habitability of temperate sub-Neptunes by detecting signature gases. See text for details.
}
\label{fig:population}
\end{figure}

The temperate sub-Neptune K2-18 b is a harbinger of the class of planets that might be habitable and exemplifies the need for a near-term method to measure the size of an H$_2$ atmosphere. The planet of $8.6\ M_{\oplus}$ and $2.6\ R_{\oplus}$ is in the habitable zone of an M dwarf star, and has a transmission spectrum (obtained by \textit{Hubble} at $1.1 - 1.7\ \mu m$) with confirmed spectral features, which indicates that the planet should host an atmosphere dominated by H$_2$ \citep{tsiaras2019water,benneke2019water}. Interior structure models showed that the planet can have a massive ($>\sim1000$ bar) H$_2$ atmosphere overlaying a rocky/Fe core and a possibly supercritical water layer, or a smaller ($<100$ bar) H$_2$ atmosphere with a water-dominated interior \citep{madhusudhan2020interior,mousis2020irradiated,nixon2021deep}. For K2-18~b, specifically, a $\sim10-100$ bar H$_2$ atmosphere overlaying a water layer would cause $>200$ bar of water to evaporate into the atmosphere, resulting in a hot steam atmosphere inconsistent with the observed transmission spectrum \citep{scheucher2020consistently}. An even smaller, $\sim1$ bar H$_2$ atmosphere would prevent this steam atmosphere and produce a liquid-water ocean (see Section~\ref{sec:model}), but this requires a very small rocky/Fe core and may be disfavored from the planet formation standpoint \citep[e.g.,][]{lee2016breeding}. However, a planet slightly more massive or smaller than K2-18~b -- such as those at the center of the $1.7-3.5\ R_{\oplus}$ planet population -- does not have this small-core difficulty to have a small atmosphere \citep{zeng2019growth,nixon2021deep}, and many such planets and planet candidates have been detected and will soon be available for transmission spectroscopy (Figure~\ref{fig:population}, panel a).

Here we propose that transit observations of temperate sub-Neptunes in the near- and mid-infrared wavelengths, which will soon commence with JWST, can detect small H$_2$ atmospheres that support liquid-water oceans and distinguish them from massive atmospheres (Figure~\ref{fig:population}, panel b). A companion paper has studied the atmospheric chemistry and spectral features of temperate planets with massive H$_2$ atmospheres \citep{hu2021photochemical}, and now we turn to temperate planets with small H$_2$ atmospheres. A recent paper might have similar intent as our work:
\cite{yu2021identify} studied the chemistry of temperate \ce{H2} atmospheres with varied surface pressures, with assumed zero flux for all species at the lower boundary. The theories of \cite{yu2021identify} may thus be more applicable to arid rocky planets without substantial volcanic outgassing, and here we instead focus on ocean planets, and address how to identify them observationally. As we will show later, a small atmosphere on a temperate sub-Neptune will have a distinctive composition because of its interaction with the ocean underneath. 

\section{Mutual exclusivity of habitability and thermochemical equilibrium} \label{method}

\begin{figure}[!htbp]
\centering
\includegraphics[width=0.45\textwidth]{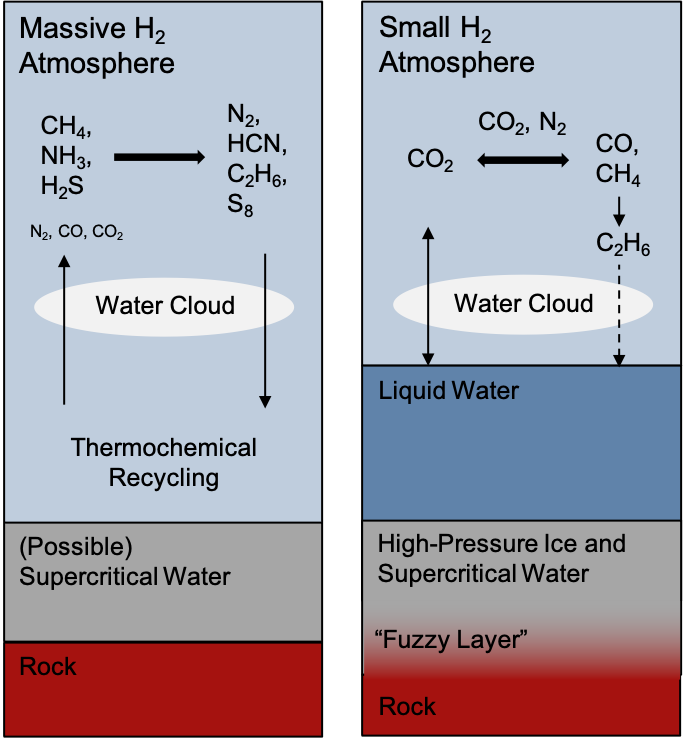}
\caption{
Interior structures of temperate H-rich exoplanets and the associated ranges of atmospheric composition. If the planet has a massive H$_2$ atmosphere, the deep atmosphere would be hot -- enabling thermochemical recycling -- but a liquid-water surface would not be possible. If the planet has a small H$_2$ atmosphere, a liquid-water surface may be possible. On these planets, the equilibrium abundance of atmospheric CO$_2$ is set by the oceanic chemistry and that of N$_2$ by atmospheric evolution.
}
\label{fig:schematic}
\end{figure}

On temperate sub-Neptunes, the condition to form a liquid-water ocean and that to achieve the thermochemical equilibrium of carbon and nitrogen molecules are mutually exclusive. The \ce{CO2}-\ce{CO}-\ce{CH4} and \ce{N2}-\ce{NH3} conversion rates are primarily a function of the temperature and to a lesser extent the pressure \citep{zahnle2014methane,tsai2018toward}, and in a temperate sub-Neptune like K2-18~b, the thermochemical equilibrium of carbon and nitrogen molecules are typically achieved at the pressure of $10^7\sim10^8$ Pa, where the temperature is $>1000$ K \citep[i.e., substantially higher than the critical point of water;][]{fortney2020beyond,yu2021identify,hu2021photochemical}. Therefore, the gas-phase thermochemical equilibrium would be achieved in the deep and hot part of a massive atmosphere, and in contrast, it would not be achieved in a small atmosphere overlying a liquid-water ocean. Instead, NH$_3$ and sulfur species would be sequestered by the ocean \citep[][and also see Section~\ref{sec:model}]{loftus2019sulfate} and the abundance of CO$_2$ would be set by the ocean chemistry (Figure~\ref{fig:schematic}, with the cosmochemical and geological constraints detailed in Appendix~\ref{sec:geo}). This fundamental difference, coupled with atmospheric photochemistry, leads to distinctive gas abundances in the observable part ($<\sim0.1$ bar) of the atmosphere.

If the planet has a massive H$_2$ atmosphere, thermochemical reactions in the deep atmosphere recycle O, C, N, S species into H$_2$O, CH$_4$, NH$_3$, and H$_2$S \citep{burrows1999chemical,heng2016analytical,woitke2020coexistence,Blain2020}. H$_2$O can form a cloud and the above-cloud H$_2$O may be partially depleted as a result \citep{morley2014water,charnay2021formation,hu2021photochemical}. Recent calculations have shown that the photodissociation of NH$_3$ in the presence of CH$_4$ leads to the formation of HCN and \ce{N2}, and that CO and CO$_2$ are produced by the photodissociation of CH$_4$ together with H$_2$O \citep{hu2021photochemical}. The photodissociation of H$_2$S leads to the formation of elemental sulfur haze \citep{hu2013photochemistry,zahnle2016photolytic}, but the haze would likely be close to the cloud deck and would not mute transmission spectral features \citep{hu2021photochemical}. These photochemical products are transported to the deep atmosphere and recycled back to CH$_4$, NH$_3$, and H$_2$S. An exception is that planets with super-solar atmospheric metallicity and appreciable internal heat may have additional \ce{CO}, \ce{CO2}, and \ce{N2} transported from the deep troposphere and incomplete recycling to \ce{NH3} \citep{fortney2020beyond,yu2021identify,hu2021photochemical}.

If the planet instead has a small atmosphere and a liquid-water ocean, the thermochemical recycling cannot occur. Instead, CO$_2$ is the preferred form of carbon in equilibrium with a massive amount of H$_2$O \citep{Hu2014B2014ApJ...784...63H,woitke2020coexistence}, and NH$_3$ is dissolved in the ocean and largely depleted from the atmosphere (see Section~\ref{sec:model}). The abundance of atmospheric CO$_2$ is controlled by the oceanic pH \citep{kitzmann2015unstable,krissansen2017constraining,kite2018habitability,isson2018reverse} and that of N$_2$ is probably a combined result of the initial endowment and atmospheric escape. A reasonable lower bound of the total mass of CO$_2$ in the H$_2$ and H$_2$O layers can be derived from the cosmochemical constraints of planetary building blocks and the partitioning between the iron core, the silicate mantle, and the water layer (Appendix~\ref{sec:geo}). Also, the ``seafloor'' of this thin-atmosphere, H$_2$O-rich sub-Neptune will not be not a sharp interface in density and composition, but instead have a finite thickness \citep{vazan2020new}. The interface will be compositionally stratified with denser material underlying less dense material, and material transport across this ``fuzzy layer'' is inhibited due to the stratification. Thus, any carbon or nitrogen added to the H$_2$ and H$_2$O envelope by planetesimal accretion late in planet growth will remain in the envelope, and will not be stirred down into the silicate layer. Meanwhile, transit observations can straightforwardly identify H$_2$-dominated atmospheres and rule out CO$_2$ or N$_2$-dominated ones only from the size of spectral features \citep{miller2008atmospheric}.

One might also consider the intermediate situation between massive atmospheres with thermochemical equilibrium and small atmospheres with liquid-water oceans, e.g., the atmospheres with a surface pressure from a few to $\sim100$ bars on K2-18~b. For many sub-Neptunes, this intermediate-atmosphere scenario would still require a massive water layer underneath to explain their mass and radius. If water was in the liquid form at the interface with the atmosphere, the evaporation of this ocean would make the atmosphere H$_2$O-dominated \citep{scheucher2020consistently}. If water is supercritical, any H$_2$ layer of intermediate mass should be well mixed with the water layer. Therefore, such an intermediate endowment of H$_2$ would most likely result in a non-H$_2$-dominated atmosphere, which is, again, distinguishable with transmission spectroscopy \citep{miller2008atmospheric}.
	
\section{Ocean Planet Models} \label{sec:model}

\begin{deluxetable*}{llll|llll}
\tablecaption{Summary of the photochemical model parameters and results.}
\label{table:result}
\tablehead{
\colhead{Model} & \colhead{Name} & \colhead{CO$_2$} & \colhead{CO flux} & \colhead{H$_2$O} & \colhead{CO} & \colhead{CH$_4$} & \colhead{C$_2$H$_6$} }
\startdata
 1 & Low-CO$_2$ & $4\times10^{-4}$ & $0$ & $2.3\times10^{-3}$ & $1.4\times10^{-5}$ & $1.5\times10^{-2}$ & $3.0\times10^{-6}$ \\
 1a & Low-CO$_2$ Variant & $4\times10^{-4}$ & $1.0\times10^9$ & $3.3\times10^{-3}$ & $2.9\times10^{-4}$ & $2.9\times10^{-2}$ & $5.1\times10^{-6}$ \\
 2 & High-CO$_2$ & $0.1$ & $0$ & $1.1\times10^{-4}$ & $9.5\times10^{-3}$ & $5.3\times10^{-2}$ & $4.0\times10^{-7}$ \\
\enddata
\tablecomments{The volume mixing ratio of CO$_2$ (as inputs) is at the lower boundary, and those of H$_2$O, CO, CH$_4$, and C$_2$H$_6$ (as results) are column-averaged in $10-10^3$ Pa. The CO flux has a unit of cm$^{-2}$ s$^{-1}$.}
\end{deluxetable*}

\begin{figure*}[!htbp]
\centering
\includegraphics[width=0.9\textwidth]{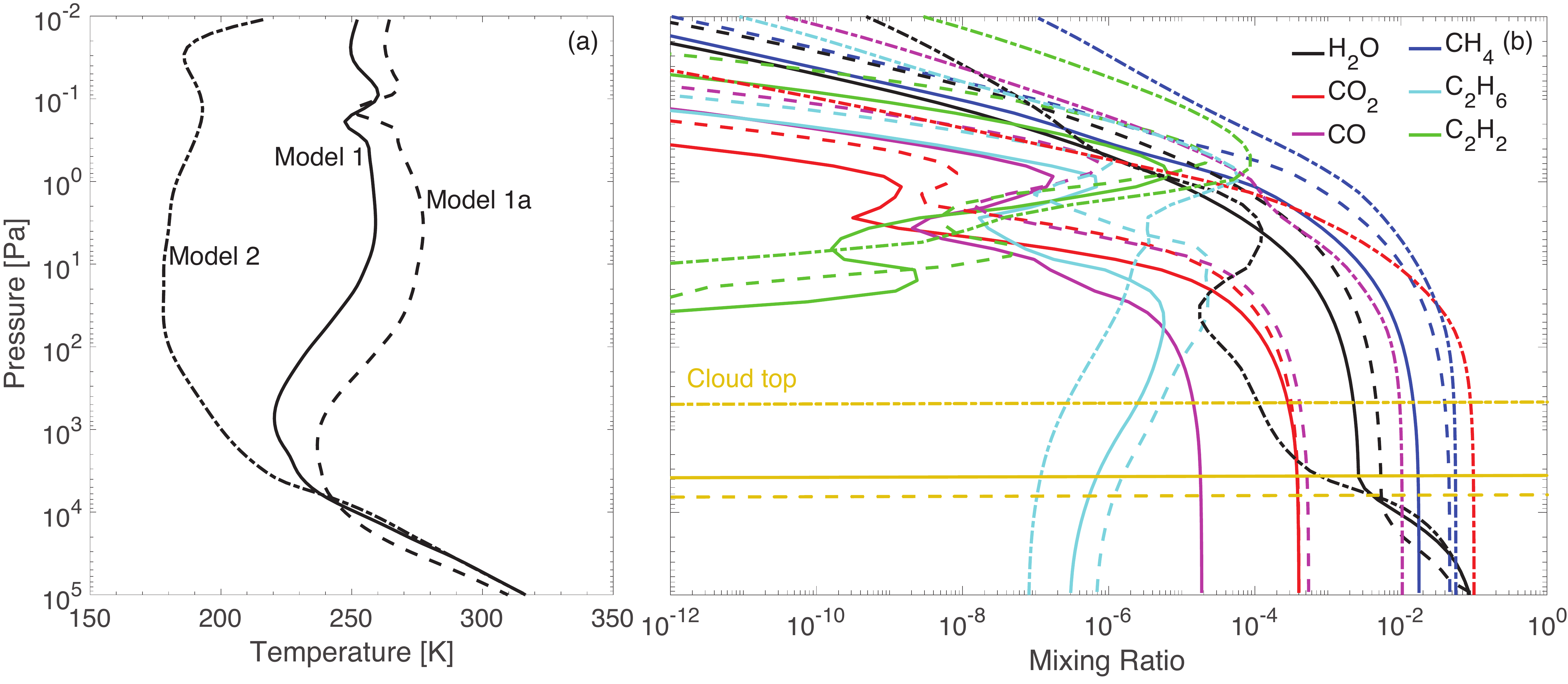}
\caption{
Modeled pressure-temperature profiles (a) and abundance profiles of main gases and photochemical products (b) in a temperate sub-Neptune like K2-18~b that has a small H$_2$ atmosphere. Solid, dashed, and dash-dot lines show the results for the low-CO$_2$ case (Model 1 in Table~\ref{table:result}), the low-CO$_2$ case with additional CO sources (Model 1a), and the high-CO$_2$ case (Model 2). For the stellar irradiance, we use $S=S_{\rm Earth}*(1-A_{\rm B})$, with a Bond albedo of $A_{\rm B}=0.3$ (similar to Earth), to account for the radiative effects clouds would have in the otherwise cloud-free climate model. The surface albedo reflects a dark ocean (0.06). The surface temperatures in these models are consistent with a liquid-water ocean. The photochemical models use the UV spectrum of the M dwarf star GJ~176 \citep{france2016muscles} \citep[similar to K2-18;][]{dos2020high}. The steady-state mixing ratio \ce{CH4} is high and those of nitrogen molecules such as \ce{NH3} and \ce{HCN} is $<10^{-12}$.}
\label{fig:result}
\end{figure*}

We have used an atmospheric photochemical model \citep{hu2012photochemistry} coupled with a radiative-convective model \citep{scheucher2020consistently} to determine the steady-state abundances of photochemical gases in small and temperate H$_2$ atmospheres, for a cosmochemically and geologically plausible range of CO$_2$ abundance, and compared the compositions and transmission spectra with the massive H$_2$ atmosphere models published in \cite{hu2021photochemical}. The massive atmosphere models explored the atmospheric metallicity of $1-100\times$solar and included possible deep-tropospheric source \ce{CO}, \ce{CO2}, and \ce{N2} and incomplete reclycing of \ce{NH3} in super-solar atmospheres.

The photochemical model includes a comprehensive reaction network for O, H, C, N, and S species (including sulfur aerosols, hydrocarbons, and the reactions important in H$_2$ atmospheres), and it has been used to study the lifetime and equilibrium abundance of potential biosignature gases in H$_2$ atmospheres \citep{seager2013biosignature}. We have updated the reaction network and tested the model with the measured photochemical gas abundance in the atmosphere of Jupiter \citep[i.e., a low-temperature H$_2$ atmosphere;][]{hu2021photochemical}. 

The pressure-temperature profiles (Figure~\ref{fig:result}) used as the basis for the photochemical model are calculated with the climate module of 1D-TERRA \citep{scheucher2020consistently}. The module uses a correlated-k approach with the random overlap method to include molecular absorption, collision-induced opacities, and the continuum of water vapor to calculate the radiative equilibrium, and the appropriate (moist or dry) adiabatic lapse rate to apply the convection adjustment. The module has been tested against the cases of Earth, Venus, and Mars, as well as with other radiative-convective and 3D climate models for modeling steam atmospheres \citep{scheucher2020consistently}.

As examples, we study H$_2$ atmospheres of 1 bar on a sub-Neptune planet that has a stellar irradiance similar to Earth and orbits around an early M star similar to K2-18. A 1-bar H$_2$ atmosphere on such a planet would likely have a surface temperature consistent with a liquid-water ocean (Figure~\ref{fig:result}). We adopt the ``ocean-planet'' interpretation of the $1.7-3.5\ R_{\oplus}$ planet population that centers at $10\ M_{\oplus}$, and  $2.5\ R_{\oplus}$ \citep{zeng2019growth,venturini2020nature}, and assume 50\% of water by mass in this study. In this interpretation, sub-Neptunes may be ocean planets with deep oceans that do not require a massive H$_2$ envelope to explain their radius, and can conceivably have moderate-size H$_2$ atmospheres. This may not be directly applicable for K2-18~b, which resides on the low-density side of the $1.7-3.5\ R_{\oplus}$ population. The specific choices of these parameters are however unimportant, because atmospheric chemistry is not sensitive to moderate changes in the surface gravity. 

CO$_2$ is the main form of carbon in thermochemical equilibrium with H$_2$O \citep{Hu2014B2014ApJ...784...63H,woitke2020coexistence}. If a liquid-water ocean exists, the partial pressure of CO$_2$ is set by atmosphere-ocean partitioning, which in turn is mainly controlled by the oceanic pH \citep{kitzmann2015unstable,krissansen2017constraining,kite2018habitability,isson2018reverse}. The pH is affected by the abundance of cations in the ocean, which come from complex water-rock reactions and dissolution of the seafloor. The rates of the processes involved are uncertain; therefore, we explore the mixing ratio of CO$_2$ from 400 ppm to 10\%, corresponding to the pCO$_2$ range from the present-day Earth to early Earth \citep{catling2017atmospheric} and including the predicted range for ocean planets \citep{kite2018habitability} that is still consistent with an H$_2$-dominated atmosphere. The $4\times10^{-4}$ bar partial pressure of \ce{CO2} in the low-CO$_2$ case, while not the absolute lower limit, is a cosmochemically and geologically reasonable lower bound of the \ce{CO2} partial pressure on an ocean planet (Appendix~\ref{sec:geo}).

The mixing ratio of N$_2$ on the modeled planet is probably set by atmospheric evolution (as opposed to the solubility equilibrium or geological recycling) and is assumed here to be 1\%. As N$_2$ only minimally participates in the chemical cycles and does not have strong spectral features in the infrared, its exact abundance is not our main concern. The photochemical model indicates that the NH$_3$ produced by photodissociation of N$_2$ in H$_2$ atmospheres has negligible mixing ratios ($<10^{-12}$).

The pressure at the water-rock boundary of a $10-M_{\oplus}$ and $2.5-R_{\oplus}$ planet is $\sim500$ GPa \citep{sotin2007mass,levi2014structure}, and this overloading pressure should suppress volcanism completely \citep{kite2009geodynamics,noack2017volcanism,kite2018habitability}. Therefore we do not include any volcanic outgassing in the standard models. As variant models, we consider the possibility of minor and intermittent sources of CO into the atmosphere. Evaporation of meteorites may provide a source of CO and CO$_2$ \citep{schaefer2017redox}, and water-rock reactions at the temperature relevant to the ``fuzzy layer'' may produce CO (and not CH$_4$ as it is thermochemically disfavored at high temperatures). The rates of these processes are unknown, but numerical experiments with the photochemical model indicate that an additional CO source of $10^{10}$ molecule cm$^{-2}$ s$^{-1}$ would lead to a steady-state abundance of CO greater than that of H$_2$, effectively resulting in a CO-dominated atmosphere. A CO source of $10^9$ molecule cm$^{-2}$ s$^{-1}$ would produce the CO-dominated atmosphere in the 10\%-CO$_2$ case but not in the 400ppm-CO$_2$ case. We therefore include a low-CO$_2$ case with the CO source of $10^9$ molecule cm$^{-2}$ s$^{-1}$ as a variant model.

Table~\ref{table:result} summarizes the input parameters and results of the photochemical models, and Figure~\ref{fig:result} shows the profiles of temperature and mixing ratios of main gases and photochemical products. CO is produced from the photodissociation of CO$_2$ and can build up to $10^{-5}$ and $10^{-2}$ mixing ratio level for the low-CO$_2$ and the high-CO$_2$ cases. OH from the photodissociation of H$_2$O destroys CO and maintains its steady-state mixing ratio. CH$_4$ is also produced photochemically and can build up to a substantial mixing ratio ($10^{-3}\sim10^{-2}$). This effectiveness in producing \ce{CH4} from \ce{CO} in temperate H$_2$ atmospheres has also been noted in \cite{yu2021identify}. Together with the high CH$_4$ mixing ratio, C$_2$H$_6$ is produced and can accumulate to a mixing ratio of $\sim10^{-6}$. C$_2$H$_2$, as expected, is short-lived and only has significant mixing ratios in the upper atmosphere. Here we have applied a deposition velocity of $10^{-5}$ cm s$^{-1}$ for C$_2$H$_6$ to account for the loss of carbon due to organic haze formation and deposition \citep{hu2012photochemistry}; removing this sink does not substantially change the results shown in Figure~\ref{fig:result}. The additional source of CO would result in moderately more CO, CH$_4$, and C$_2$H$_6$ in the atmosphere (Model 1a in Table~\ref{table:result} and Figure~\ref{fig:result}). The photochemical CO and CH$_4$ can build up to the mixing ratio levels that cause significant features in the planet's transmission spectrum (Section~\ref{sec:spec}).

Before closing this section, we address whether NH$_3$ can be produced substantially by water-rock reactions and then emitted into the atmosphere. Hydrothermal systems on early Earth may produce \ce{NH3} from the reduction of nitrite and nitrate \citep{summers1993prebiotic,summers2005ammonia}. On a planet with an \ce{H2}-dominated atmosphere, however, atmospheric production of the oxidized nitrogen including nitrite and nitrate should be very limited. Moreover, the storage capability of \ce{NH3} by the ocean is vast and limits the emission into the atmosphere. At the pH value of 8 (a lower pH would further favor the partitioning of \ce{NH3} in the ocean), $10^{-6}$ bar of atmospheric \ce{NH3} requires a dissolved ammonium concentration of $10^{-3}$ mol/L in equilibrium \citep{seinfeld2016atmospheric}. The mass of NH$_3$ in the atmosphere and ocean is then $\sim10^{-5}$ of the planetary mass. This would only be possible if much of the planet's rocky core begins with a volatile composition similar to carbonaceous chondrites, and most of this nitrogen is partitioned into the atmosphere and ocean as NH$_3$ \citep{marty2016origins}, which is highly unlikely as \ce{N2} is thermochemically favored. Therefore, the concentration of dissolved \ce{NH3} should be small and so is the atmospheric \ce{NH3} on a planet with a massive ocean.

\section{Spectral Characterization} \label{sec:spec}

\begin{figure*}[!htbp]
\centering
\includegraphics[width=0.9\textwidth]{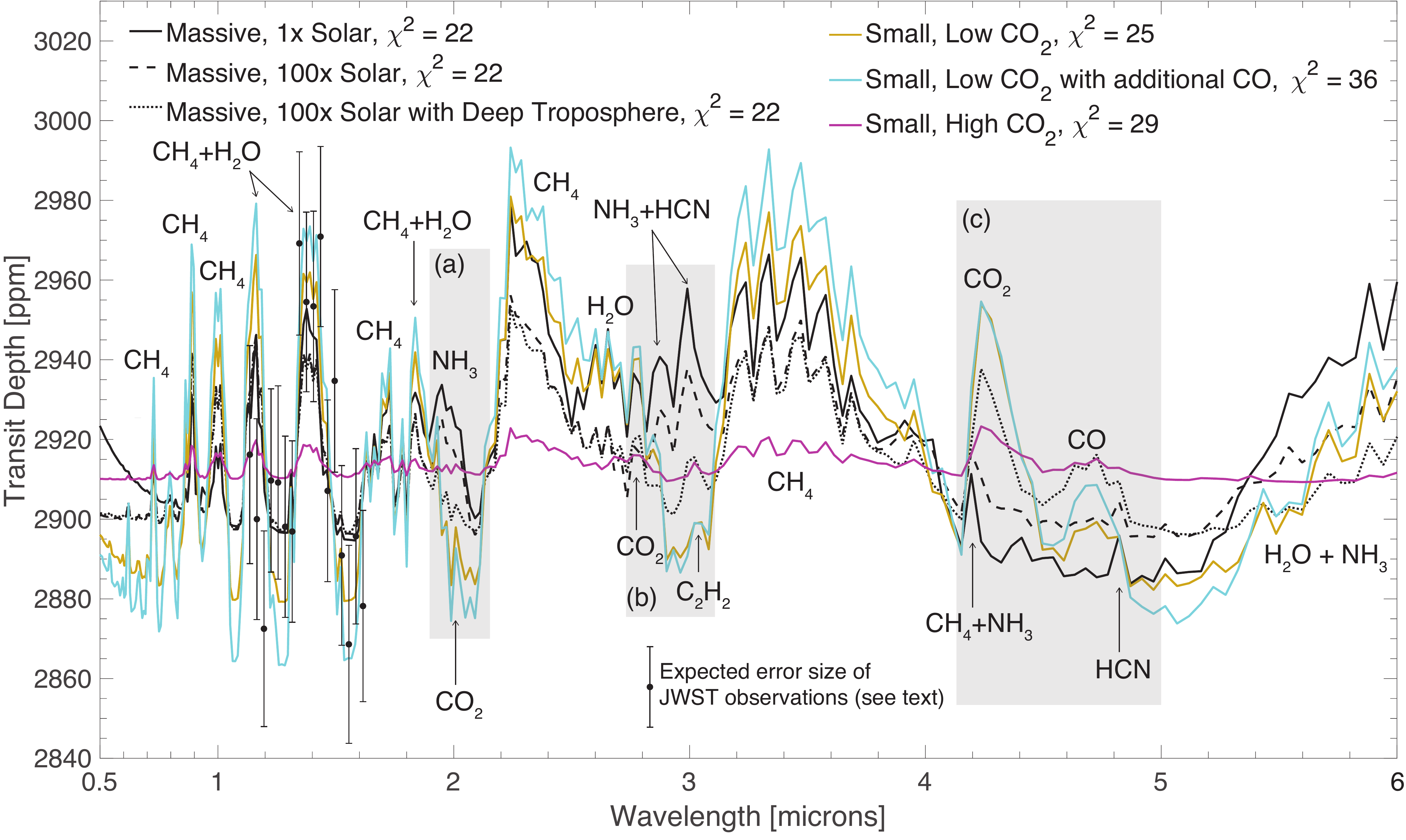}
\caption{
Modeled transmission spectrum of temperate sub-Neptune planets of M dwarf stars, using K2-18~b as an example and comparing with the planet's transit depth observed by \textit{Hubble} \citep{benneke2019water}. The massive-H$_2$-atmosphere models (black lines) and the small-H$_2$-atmosphere models (colored lines) differ in three spectral regions: in (a) and (b), the massive-atmosphere models have absorption features of NH$_3$ and HCN, while the small-atmosphere models do not; in (c), the small-atmosphere models with a low mixing ratio of CO$_2$ (400 ppm) have prominent features of CO$_2$ and CO, while the massive-atmosphere models only have small features of NH$_3$ and HCN. The $100\times$solar massive atmosphere with deep-tropospheric source and sink may have subdued NH$_3$ and HCN features and prominent CO$_2$ and CO features. The small-atmosphere models with a high mixing ratio of CO$_2$ (10\%) has a high mean molecular weight ($\sim6$) and a high cloud top (Figure~\ref{fig:result}) and thus muted spectral features.
}
\label{fig:spec}
\end{figure*}

Figure~\ref{fig:spec} compares the expected spectra for the massive-atmosphere scenarios and the small-atmosphere scenarios. For K2-18~b, the massive-atmosphere models with $1-100\times$solar metallicity and the small-atmosphere models with a low mixing ratio of CO$_2$ (400 ppm) provide good fits to the transmission spectrum measured by \textit{Hubble}. 

Measuring the transmission spectra in an expanded wavelength range of $1-5\ \mu$m will distinguish the small atmospheres from massive ones. Using K2-18~b as an example for temperate sub-Neptunes, we see that the massive-atmosphere models and the small-atmosphere models, while having differences within each group, can be distinguished using the spectral regions of $1.9-2.1$, $2.7-3.1$, and $4.1-5.0\ \mu$m (the shaded areas a, b, and c in Figure~\ref{fig:spec}). Both the massive-atmosphere and small-atmosphere models show spectral features of H$_2$O and CH$_4$, and so observing these two gases alone is unlikely to separate the massive versus small scenarios. 

At $1.9-2.1$ and $2.7-3.1\ \mu$m, the transmission spectra show NH$_3$ and HCN absorption in massive atmospheres but not in small atmospheres. If the $100\times$solar massive atmosphere has incomplete \ce{NH3} recycling in the deep troposphere, it will have much weaker NH$_3$ and HCN features in these spectral regions. The transmission spectra of small atmospheres show small CO$_2$ features at $\sim2.0$ and $\sim2.75\ \mu$m, but the feature at $\sim2.75\ \mu$m is combined with a part of the H$_2$O feature with similar strength. The transmission spectra of small atmospheres also show a small C$_2$H$_2$ feature at $\sim3.05\ \mu$m, and given enough precision, it might be distinguishable with the HCN feature at $\sim3.0\ \mu$m.

At $4.1-5.0\ \mu$m, the transmission spectra of small atmospheres (the low-CO$_2$ cases) have prominent features of CO$_2$ and CO, while the spectra of massive atmospheres have weak features of NH$_3$ and HCN. If the $100\times$solar massive atmosphere has CO and CO$_2$ transported from the deep troposphere, it can have prominent spectral features of CO$_2$ and CO in this region as well.

From the above, we see that the $100\times$solar massive atmosphere with deep-tropospheric effects may resemble a small atmosphere in their transmission spectra (Figure~\ref{fig:spec}), i.e., the lack of \ce{NH3} or \ce{HCN} and the prominence of \ce{CO2} and \ce{CO}. Would this potential ``false positive'' be avoidable? The answer may be yes given enough precision and spectral resolution. First, the spectrum of the massive atmosphere with deep-tropospheric effects still has weak spectral features of HCN, while none of the small atmospheres does. Second, the massive atmosphere has CO$_2$/CO$<\sim0.1$, because CO always dominates over CO$_2$ in the deep H$_2$ troposphere of a temperate planet, and photochemical processes driven by an M dwarf star do not significantly raise the CO$_2$ mixing ratio in the observable part of the atmosphere \citep{hu2021photochemical}. In contrast, the small atmospheres typically have CO$_2$/CO$\geq1$ (Table~\ref{table:result}). In the more likely scenario without any volcanic outgassing, CO$_2$/CO$\sim10$, because CO is produced photochemically from CO$_2$.  Therefore, by measuring the abundance of CO and CO$_2$ independently, one could tell whether they are sourced from the deep troposphere.

Furthermore, a massive atmosphere with $\gg100\times$ solar metallicity will have a mean molecular weight much higher than that of an H$_2$ atmosphere and is thus also distinguishable by transmission spectroscopy. 

With moderate time investment (i.e., $<100$ hours), JWST will provide the sensitivity to detect the signature gases aforementioned and distinguish massive versus small atmospheres on planets like K2-18~b. As an example, we have used PandExo \citep{batalha2017pandexo} to simulate the expected photometric precision using JWST's NIRSpec instrument. If combining two transit observations with NIRSpec’s G235H grating and four transits with the G395H grating, the overall photometric precision would be $\sim20$ ppm per spectral element at a resolution of $R=100$ in both channels that cover a wavelength range of $1.7-5.2\ \mu$m. These observations would distinguish the small-atmosphere scenarios versus the massive-atmosphere scenarios in Figure~\ref{fig:spec} with high confidence. 

\begin{figure*}[!htbp]
\centering
\includegraphics[width=0.9\textwidth]{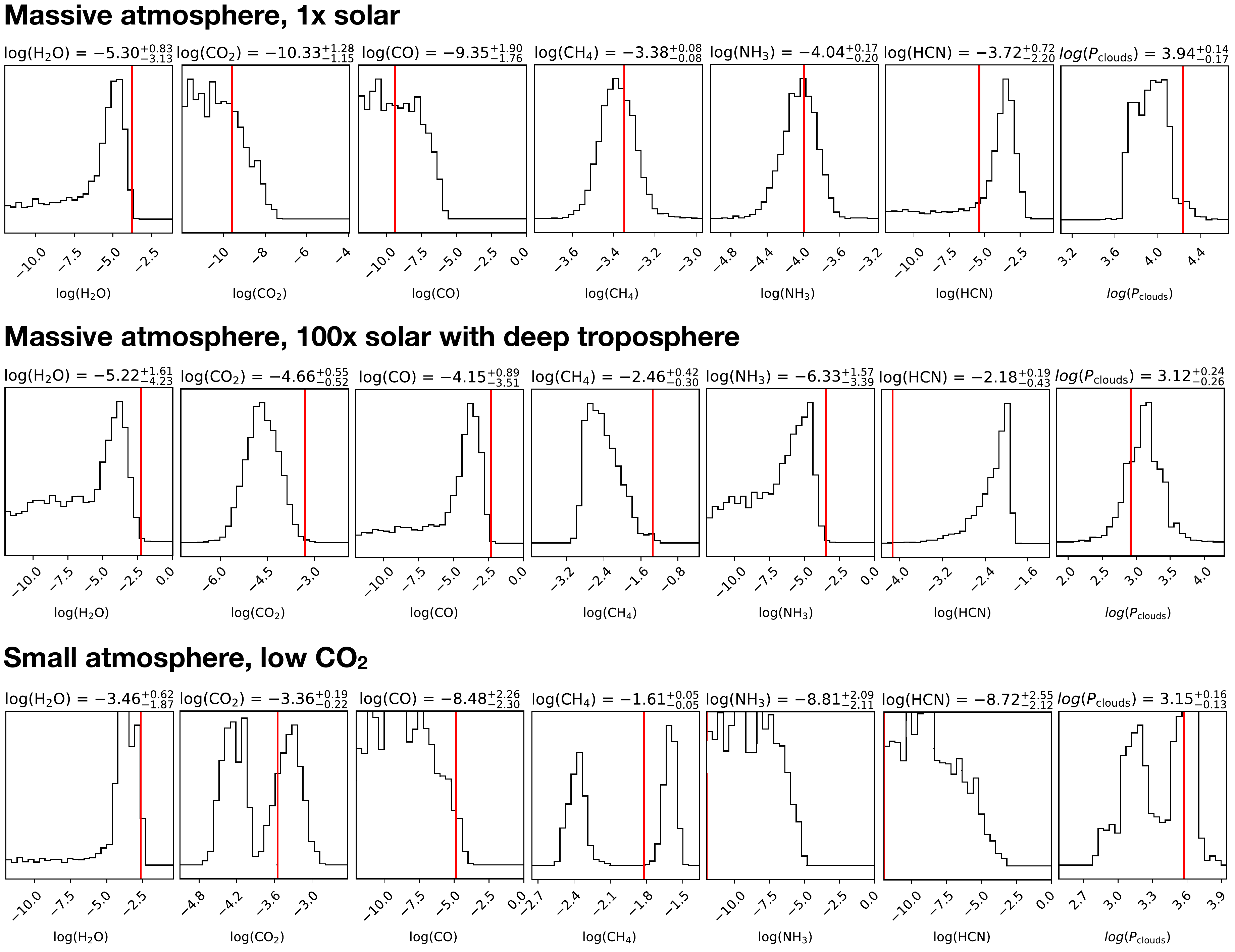}
\caption{
Retrieved posterior distributions of the abundances of the main chemical compounds and the cloud pressure in example massive-atmosphere and small-atmosphere scenarios. The input transmission spectra are calculated by Tau-REx using the atmospheric composition in Figure~\ref{fig:result} and the expected uncertainties are calculated using PandExo \citep{batalha2017pandexo}, assuming to combine two transits of K2-18~b with NIRSpec/G235H and four transits with NIRSpec/G395H. The vertical red lines show the input value of the parameter, and the quantities on the top of each panel show the median and $1\sigma$ values summarized from the posterior. The cloud pressure ($P_{\rm clouds}$) has a unit of Pa. A detailed characterization of the atmosphere of K2-18 b, including distinguishing a small atmosphere versus a massive one and measuring the abundances of \ce{H2O}, \ce{CH4}, \ce{NH3}, \ce{HCN}, \ce{CO2}, and \ce{CO}, will be achievable with moderate time investment of JWST.
}
\label{fig:posterior}
\end{figure*}

Additionally, we have performed spectral retrievals based on simulated observations using Tau-REx \citep{waldmann2015tau}. We find that the mixing ratio of NH$_3$ and HCN and the lack of CO$_2$ or CO in the solar-abundance massive atmosphere would be usefully constrained (Figure~\ref{fig:posterior}). For the $100\times$solar atmosphere, the CO$_2$ and CO transported from the deep troposphere would be identified, and the posteriors suggest that CO is likely more abundant than CO$_2$. The reduction in the mixing ratios of NH$_3$ and HCN due to incomplete recycling could also be seen in the retrieval, although the constraints on the mixing ratio of HCN is not accurate. For the small atmosphere, the retrieval yields degenerate solutions and thus double peaks in some posterior distributions. Despite this, it is clear from the posteriors that the atmosphere likely has high mixing ratios of both CO$_2$ and CH$_4$, has more CO$_2$ than CO, and has very little NH$_3$ or HCN (Figure~\ref{fig:posterior}). In addition to JWST, the dedicated exoplanet atmosphere characterization mission ARIEL could also provide the sensitivity to detect these gases with more repeated transit observations \citep{changeat2020disentangling}. This example shows that transit observations in the coming years can tell apart temperate sub-Neptunes with small H$_2$ atmospheres versus the planets with massive atmospheres and reveal their distinct atmospheric composition.

\section{Discussion and Conclusion} \label{sec:discussion}

Taken together, the results presented above identify a near-term path to detect small H$_2$ atmospheres that can be consistent with liquid-water oceans on temperate exoplanets. H$_2$ atmospheres are probably the only type of temperate atmospheres readily within the reach of JWST and ARIEL for detailed studies, since to characterize a heavier H$_2$O, N$_2$, or CO$_2$ atmosphere will require co-adding a few tens transits -- something not impossible but probably very hard \citep{belu2011primary,krissansen2018detectability,wunderlich2019detectability,pidhorodetska2020detectability,gialluca2021characterizing}. The mass of the H$_2$ atmospheres -- a parameter that is not directly measured by transits but critical for habitability if the planet is moderately irradiated -- can be inferred from transmission spectra via the signature gases that indicate solubility equilibria versus gas-phase thermochemical recycling. The biggest uncertainty is probably the temperature at the $100\sim1000$-bar pressure level in the massive-atmosphere scenarios, which may be affected by ad hot heating mechanisms such as tidal heating. Detailed models of the interior temperature and mixing may further constrain this uncertainty \citep{fortney2020beyond,yu2021identify}. Based on the range of the parameter space explored, we suggest that the sensitivity of multiple gases provided by future observatories' expanded wavelength coverage over \textit{Hubble} would enable broad categorization of small versus massive atmospheres, summarized as a roadmap in Figure~\ref{fig:population}, panel b. 

How many sub-Neptunes could we expect to be ocean planets in the first place? The current population statistics of planets provide indirect evidence that most sub-Neptunes are not ocean planets \citep{fulton2018california,owen2017evaporation,jin2018compositional}, but most known planets are hotter than planets that can be habitable. Even if the current statistics apply to temperate planets, there is plenty of room to have 10-20\% of sub-Neptunes be ocean planets, which will still be a lot of planets. Also, some planets in or just below the ``radius valley'' may be sub-Neptunes that have evolved into ocean planets \citep{kite2021water} and retained some residual H$_2$ \citep{misener2021cool}. For these reasons, this possibility of an ocean planet shrouded by a small H$_2$ atmosphere should motivate detailed observations of temperate planets with radius from near the ``radius valley'' ($\sim1.7\ R_{\oplus}$) to the main sub-Neptune population ($\sim2.5\ R_{\oplus}$). If some of the temperate planets in the aforementioned group have small H$_2$ atmospheres, their relative ease for transit observations would significantly enhance the prospect of detecting and characterizing potentially habitable exoplanets within the next decade.
	
\section*{Acknowledgments}
The authors thank helpful discussions with Fabrice Gaillard and Sukrit Ranjan. RH conceived and designed the study, simulated the photochemical models, interpreted the results, and wrote the manuscript. MD performed the JWST observation simulations and atmospheric retrievals. MS computed the pressure-temperature profiles. EK derived the cosmochemical and geological lower bounds for the carbon content. SS contributed interior structure models and insights. HR oversaw the development of the radiative-convective model used in the study. All authors commented on the overall narrative of the paper. The raw data that are used to generate the figures in this paper are available from the corresponding author upon reasonable request. This work was supported in part by NASA Exoplanets Research Program grant \#80NM0018F0612. The research was carried out at the Jet Propulsion Laboratory, California Institute of Technology, under a contract with the National Aeronautics and Space Administration.
	
	{	\small
		\bibliographystyle{apj}
		\bibliography{bib.bib}
	}
	
\appendix

\section{Reasonable lower bound of \ce{CO2}} \label{sec:geo}
	
Is the 400-ppm \ce{CO2}, or $4\times10^{-4}$ bar partial pressure in a 1-bar atmosphere, a reasonable lower bound of the \ce{CO2} partial pressure on an ocean planet? We consider this question from a cosmochemical and geochemical perspective. Assuming equilibrium (during planet formation) between a Fe-core, a silicate mantle, and a well-mixed supercritical volatile envelope, the partitioning of C mass between reservoirs is described by
\begin{equation}
{\rm C_{total} = C_{core} + C_{silicate} + C_{envelope},}
\label{eq1}
\end{equation}
where all reservoir masses are in kg, and
\begin{equation}
{\rm C_{core}/M_{core} = D_C (C_{silicate}/M_{silicate}),} 
\label{eq2}
\end{equation}
where ${\rm D_C}$ is a dimensionless partition coefficient, ${\rm M_{core}}$ (kg) is the mass of the Fe-dominated core, and ${\rm M_{silicate}}$ (kg) is the mass of the silicate mantle (molten during planet formation). For the partitioning between the envelope and the silicate mantle,
\begin{equation}
{\rm C_{envelope}k(g_{esi}/A_{esi})(\mu_{avg}/\mu_C)s_C = C_{silicate}/M_{silicate},} 
\label{eq3}
\end{equation}
where ${\rm k}$ is a stochiometric correction from C mass to the mass of the C-bearing species in the envelope (i.e., $44/12\sim3.7$ for CO$_2$), ${\rm g_{esi}}$ is gravitational acceleration at the envelope-silicate interface in m s$^{-2}$, ${\rm A_{esi}}$ is the area of the envelope-silicate interface in m$^2$, ${\rm \mu_{avg}}$ is the average molecular weight of the envelope (in Da), ${\rm \mu_C}$ is the molecular weight of the C-bearing species (in Da), and ${\rm s_C}$ is the solubility of the C-bearing species (in Pa$^{-1}$). Here we have assumed that the molten silicate layer is well-stirred.

Supposing ${\rm M_{core}/M_{silicate} \sim 0.5}$ (like Earth) and ${\rm D_C \sim 10^3}$ \citep{dasgupta2013ingassing}, then ${\rm C_{core}/C_{silicate} \sim 500}$. If ${\rm C_{silicate}/M_{silicate} \sim 50}$ ppm then ${\rm C_{core}/M_{core} \sim 2.5}$ wt\%, or ${\rm C_{total}/(M_{core}+M_{silicate}) \sim 1}$ wt\%, which is a reasonable lower bound for the primordial carbon endowment (see below). For ${\rm s_C = 0.55}$ ppm/Mpa \citep{dasgupta2019origin}, the envelope partial pressure of the C species ${\rm (= C_{envelope}k (g_{esi} / A_{esi} ) ( \mu_{avg} / \mu_C ))}$ is $\sim10^3$ bars. For a ${\rm 5-M_{\oplus}}$ and ${\rm 1.5-R_{\oplus}}$ core+mantle \citep{zeng2019growth} that defines the envelope-silicate boundary, and ${\rm \mu_{avg} / \mu_C} = 0.4$ (appropriate for CO$_2$ in a H$_2$O-dominated supercritical layer during planet formation), the CO$_2$ mass in the envelope is 0.2\% of an Earth mass. This estimate shows that even though most C is in the core, still-significant reservoirs of C exist both in the silicate and in the envelope \citep{dasgupta2019origin,bergin2015tracing,keppler2019graphite,hirschmann2016constraints}. Recent indications that the partition coefficient ${\rm D_C}$ is $\ll10^3$ at the pressures and temperatures that are relevant for assembly of sub-Neptunes \citep{fischer2020carbon} would imply even more envelope C enrichment.

Following the formation of the liquid-water ocean, almost all of the \ce{CO2} will be dissolved in the ocean. For a ${\rm 5-M_{\oplus}}$ water layer, the CO$_2$ mass in the envelope estimated above corresponds to a concentration of $\sim0.01$ mol/L of dissolved CO$_2$. Here we have also assumed that the ocean is well-stirred. A higher oceanic pH leads to more effective dissolution and less CO$_2$ in the atmosphere. As an extreme, if cations are leached from the silicate and not charge-balanced by chloride ions, then an ocean composition with a pH of 9 -- 10 (``a soda lake'') will result \citep{kempe1985early}. Using the equilibrium constant of carbonate and bicarbonate dissociation \citep{seinfeld2016atmospheric}, the CO$_2$ partial pressure in equilibrium with this ocean would be $5\times10^{-5}\sim7\times10^{-4}$ bar, which is consistent with the assumed lower bound.

The partition coefficient gives the ratios of concentration of a species in the Fe-dominated core to the concentration of the same species in the silicate mantle. Therefore doubling the total amount of C in the core+mantle will double the concentration in the magma. What is the whole-planet C content? In principle, a planet can form without accreting volatiles. However, a thin-atmosphere sub-Neptune must have a thick volatile (H$_2$O) layer in order to match density data. It is very likely that a world that forms with 10s of wt\% H$_2$O will also accrete abundant C. We develop this point in more detail in the following paragraph.

At $T_{\rm eff}\sim300$ K, the minimum liquid water content to explain most sub-Neptune masses and radii is $>\sim50$ wt\% even if there is no Fe-metal core \citep{mousis2020irradiated}. This is more H$_2$O than can possibly be produced by hydrogen-magma reactions \citep{kite2021water}, and instead implies a contribution of planet building blocks from the temperature range beyond the water ice snowline. This is a zone where (in the Solar System), abundant refractory carbon is found. Specifically, the carbon content of primitive chondrite meteorites (CI and CM type) is 2-6 wt\% \citep{pearson2006carbon}. Although we do not fully understand the origin of this refractory carbon, proposed mechanisms for forming this refractory carbon would also apply to exoplanetary systems \citep{bergin2014exploring}. Therefore we assume a planet bulk composition of ${\rm (2-6\ wt\%)\times(1 - x)}$ carbon, where ${\rm x}$ is the H$_2$O mass fraction, and the remainder of the planet's building blocks are assumed to have a C content similar to that of primitive chondrites. This is a conservative lower limit on bulk C content for a thin-H$_2$-atmosphere sub-Neptune, for the following two reasons. (i) It considers only refractory C, not C ices (e.g., CO$_2$ ice) which could be important in the case of whole-planet migration. (ii) Some primitive bodies in the Solar System appear to be more C-rich than the most primitive chondrite meteorites; for example, the surface of the dwarf planet Ceres may contain 20 wt\% C \citep{marchi2019aqueously}. These large bulk C contents map to substantial envelope C contents (Equations \ref{eq1}-\ref{eq3}). As such, the $4\times10^{-4}$ bar partial pressure of \ce{CO2}, while not the absolute lower limit, is a cosmochemically and geologically reasonable lower bound of the \ce{CO2} partial pressure on an ocean planet.
	
\end{document}